# Field-induced quantum critical point in the pressure-induced superconductor CeRhIn$_5$

Tuson Park [*,1,2], Y. Tokiwa[2], F. Ronning[2], H. Lee[2], E. D. Bauer[2], R. Movshovich[2], J. D. Thompson[2]

[1] Department of Physics, Sungkyunkwan University, Suwon 440-746, Republic of Korea
[2] Condensed Matter and Thermal Physics, Los Alamos National Laboratory, Los Alamos, NM 87545, USA

When subjected to pressure, the prototypical heavy-fermion antiferromagnet CeRhIn$_5$ becomes superconducting, forming a broad dome of superconductivity centered around 2.35 GPa (=$P2$) with maximal $T_c$ of 2.3 K. Above the superconducting dome, the normal state shows strange metallic behaviours, including a divergence in the specific heat and a sub-$T$-linear electrical resistivity. The discovery of a field-induced magnetic phase that coexists with superconductivity for a range of pressures $P \leq P2$ has been interpreted as evidence for a quantum phase transition, which could explain the non-Fermi liquid behavior observed in the normal state. Here we report electrical resistivity measurements of CeRhIn$_5$ under magnetic field at $P2$, where the resistivity is sub-$T$-linear for temperatures above $T_c$ (or $T_{FL}$) and a $T^2$-coefficient $A$ found below $T_{FL}$ diverges as $H_{c2}$ is approached. These results are similar to the field-induced quantum critical compound CeCoIn$_5$ and confirm the presence of a quantum critical point in the pressure-induced superconductor CeRhIn$_5$.

## 1. Introduction

Proximity of a magnetic phase to superconductivity, commonly observed in unconventional superconductors, suggests that fluctuating spins play a role in producing unconventional superconductivity in a way analogous to phonons in conventional superconductors. Tuning external parameters, such as chemical substitution or pressure, suppresses the magnetic order and a dome of unconventional superconductivity (SC) appears in the vicinity of a quantum critical point where the magnetic transition temperature extrapolates to zero and spin fluctuations become singular [1]. The superconducting dome, however, inherently veils the presence of a quantum critical point, making it difficult not only to identify the critical point but also to understand the interplay between them [2]. Magnetic field, another tuning parameter, has been shown to be effective in lifting this veil, revealing a magnetic quantum phase transition from d-wave superconductivity to a coexisting phase of magnetism and superconductivity both in high-$T_c$ cuprates [3] and in heavy-fermion superconductors [4]. A growing body of experiments suggest commonalities among different classes of unconventional superconductors [5] and point toward an universal mechanism from which complex phenomena are derived in combination with material characteristics.

CeRhIn$_5$, a member of the heavy-fermion family CeMIn$_5$ (M=Co, Rh, Ir), is an antiferromagnet at ambient pressure and becomes superconducting with applying pressure [6], where SC properties are consistent with a d-wave gap with line nodes [7, 8]. The normal state from which superconductivity arises deviates from classic Landau-Fermi liquid behaviour: the dependence on temperature of electrical resistivity is distinctly not $T^2$ and the specific heat coefficient $C/T$ does not saturate, rather diverges with decreasing temperature [4, 9]. Below a critical pressure $P1$, where the antiferromagnetic (AFM) transition temperature $T_N$ becomes equal to $T_c$, local magnetism and superconductivity coexist on a microscopic scale [10]. Unlike chemical substitution, as in high-$T_c$ and other heavy fermion superconductors, pressure tunes the electronic configuration and thus the ground state of $CeRhIn_5$ without incurring additional disorder, providing an opportunity to study intrinsic properties driven purely by strong correlation effects. Here we report electrical resistivity of $CeRhIn_5$ at 2.35 GPa (=$P2$), where non-Fermi-liquid behaviour appears down to the lowest experimental temperature as superconductivity is suppressed by an applied field. For fields greater than $H_{c2}(0)$, a $T^2$-resistivity develops at low temperatures with a coefficient $A$ that diverges as $H_{c2}(0)$ is approached from above, indicating that the field-induced magnetic quantum critical point is close to $H_{c2}$ at this pressure.

---

* Corresponding author: e-mail tp8701@skku.edu, Phone: +82-31-299-4543, Fax: +82-31-290-7055

## 2. Experimental methods

Plate-like single crystals of $CeRhIn_5$ were prepared by using In-flux method, which is described elsewhere [6]. A conventional four-probe technique was used to measure electrical resistivity of $CeRhIn_5$ for current flowing both parallel $(\rho_c)$ and perpendicular $(\rho_{ab})$ to the tetragonal c-axis. At ambient pressure $\rho_{ab}$ is 23.08 and 0.0287 $\Omega\cdot$cm at 300 and 1.0 K, respectively, giving a residual resistivity ratio (RRR), $\rho_{ab}(300K)/\rho_{ab}(1K)$, of more than 800 and indicating that the single crystals for this study are of exceptional high quality, almost free from disorder. Resistivity measurements under pressure and magnetic field were performed with a hybrid clamp-type pressure cell down to 70 mK. Good hydrostaticity was achieved using silicone fluid as a pressure medium and precise values of pressure at low temperatures were determined by inductively measuring the shift of Sn's superconducting transition temperature [11].

## 3. Results and Discussion

Previously, we have discussed the temperature dependence of the electrical resistivity of $CeRhIn_5$ under pressure and at zero magnetic field [9]. In the high pressure limit ($P >> P2$) where both superconductivity and magnetism are suppressed, the resistivity recovers a Landau-Fermi-liquid $T^2$ dependence. In contrast, when $CeRhIn_5$ is subject to optimal pressure ($P2$) where $T_c$ is highest, the resistivity shows a sub-$T$-linear behaviour over an extensive temperature range. Even though there is no magnetic order at $P2$ in the absence of a magnetic field, this sub-$T$-linear resistivity suggests that a magnetic quantum critical point at $T=0$ may be hidden by the dome of unconventional superconductivity. This interpretation of a hidden quantum critical point is supported by a divergence of $C/T$ with decreasing temperature, where $C$ is the specific heat of $CeRhIn_5$ [4]. Anisotropy between the Ce-In intra-planar and inter-planar resistivities reveals the nature of the proposed quantum criticality. In the high pressure limit, Kondo coherence effects dominate, and the transport anisotropy ($\rho_c/\rho_{ab}$) increases with decreasing temperature up to a factor of 5 in the Landau-Fermi-liquid regime. In the quantum critical regime, in contrast, resistivity anisotropy does not change with decreasing temperature, which reflects the local nature of quantum critical fluctuations. In this communication, we focus on transport properties at 2.35 GPa, the optimal pressure, under magnetic field, which confirms the presence of a field-induced quantum critical point.

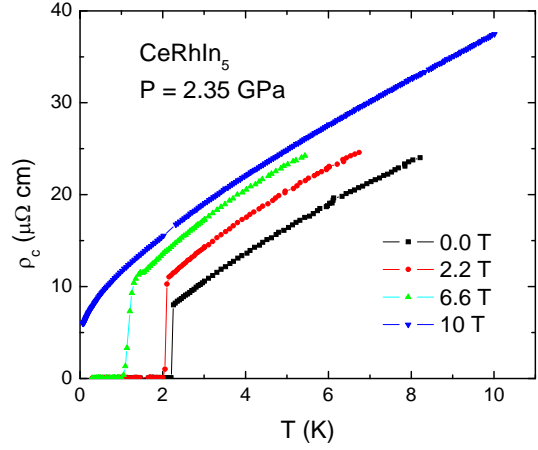

**Figure 1** C-axis resistivity of CeRhIn$_5$ at 2.35 GPa. The c-axis resistivity is plotted as a function of temperature for 0, 2.2, 6.6, and 10 Tesla.

The inter-planar electrical resistivity $\rho_c$ of CeRhIn$_5$ at 2.35 GPa is displayed as a function of temperature in Fig. 1 on a linear scale. At zero magnetic field, it drops to zero resistance at 2.2 K due to the SC transition. With increasing magnetic field, the SC transition temperature $T_c$ is suppressed and the zero-temperature upper critical field is located near 9.3 Tesla. The initial slope near $T_c$ is -11 T/K, which implies an orbital depairing field $H_{orb}(0)$ of 17.6 Tesla in the clean limit [12]. With the assumption that the Pauli limiting field $H_p$ is equal to $H_{c2}(0)$, the Maki parameter $\alpha = 2^{1/2} H_{orb}/H_p$ is approximately 2.6, which is larger than the minimum value of 1.8 that is required for an FFLO (Fulde-Ferrell-Larkin-Ovchinnikov) state [13]. Instead of saturating at the lowest temperatures, the upper critical field slightly increases (see Fig. 3a). Further specific heat or spin-lattice relaxation rate measurements under pressure at low temperatures and high fields may shed light on whether there is a realization of an FFLO state in CeRhIn$_5$.

In Fig. 2a, $\Delta\rho_c/A$ of CeRhIn$_5$ at 2.35 GPa is plotted against $T^n$ for $H=$ 0 (squares) and 10 T (down triangles), where $\Delta\rho_c = \rho_c - \rho_0 = AT^n$. At 0 T, the resistivity follows a simple power law of $T^{0.83}$ over a large temperature range from $T_c$ (~2.3 K) to $T_{max}/2$ (~17 K), where $T_{max}$ is the resistivity maximum temperature. Applying a magnetic field not only suppresses $T_c$ but also extends the temperature range toward low temperatures where the same power-law behaviour remains valid (not shown). At 10 T, which is slightly higher than $H_{c2}(0)$, the resistivity shows a sub-$T$-linear behaviour down to 400 mK and a crossover to Landau-Fermi-liquid $T^2$ dependence below 200 mK. Figure 2b shows the dependence on magnetic field of the residual resistivity $\rho_0$ (on the left ordinate) and the temperature exponent $n$ (on the right ordinate). The exponent $n$, which is centred around the value of 2/3, has a non-trivial field dependence, ranging from 0.83 at 0 T to 0.55 at 5 T. The residual resistivity, which is higher at $P2$ than at $P=0$, still is relatively small at low fields but starts to increase rapidly above 7 T, showing a maximum near $H_{c2}$ (~9.3 T). This sharp increase could be ascribed to enhanced impurity scattering due to quantum fluctuations. Above 13 T, $\rho_0$ increases again, which may be associated with a large magneto resistance at high magnetic fields.

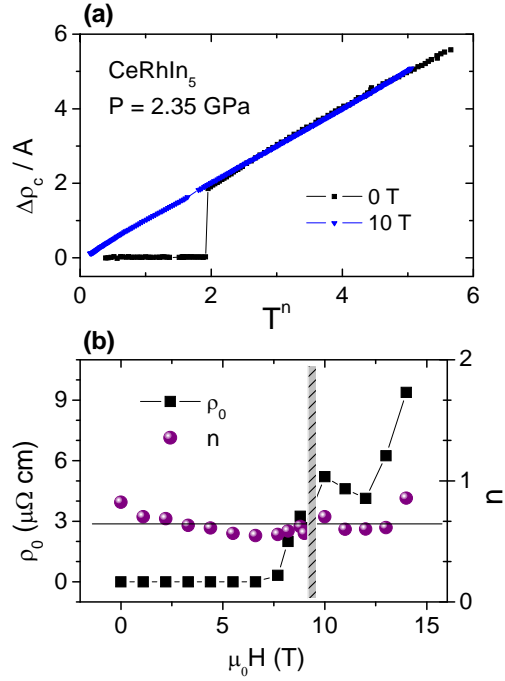

**Figure 2** (a) Normalized resistivity ($\Delta\rho_c/A$) as a function of $T^n$, where $\Delta\rho_c = \rho_c - \rho_0 = AT^n$ at 0 (squares) and 10 Tesla (down triangles). (b) Magnetic field dependence of the residual resistivity $\rho_0$ and the exponent $n$, given by left and right ordinates respectively. The hashed mark represents the upper critical field (9.3 T) at this pressure.

The sub-$T$-linear dependence of the resistivity of CeRhIn$_5$ is different from that of non-Fermi-liquid behaviour observed in conventional quantum critical metals, where $T^{3/2}$ and $T$-linear behaviour are predicted for 3- and 2-dimensional antiferromagnetic critical fluctuations, respectively. A preliminary calculation by Zhu *et al.* shows that the scattering rate can be a sub-$T$-linear when there is no momentum dependence in the scattering, which could be realized in a local type of quantum criticality [14]. An alternative suggestion could be that the non-trivial temperature exponent of the resistivity arises from multi-critical fluctuations of magnetic as well as valence degrees of freedom, which occur simultaneously at the quantum critical point of CeRhIn$_5$, which has yet to be supported by theory. Yet another possibility is the criticality is of a Kondo-breakdown type, which in a 2-dimensional system gives a resistivity proportional to $T^{2/3}$ and $C/T$ diverging as $T^{-1/3}$ [15]. More experimental as well as theoretical studies are needed to understand the anomalous power-law resistivity.

Figure 3b is an expanded view of the low-temperature resistivity at 2.35 GPa for $H=$ 10 and 14 T, which shows a $T^2$ Landau-Fermi-liquid behaviour. The temperature below which the resistivity is $T^2$, $T_{FL}$, is marked by arrows and increases with increasing magnetic field. The solid lines in Fig. 3b are least-squares fits with $\rho_c = \rho_0 + AT^2$. Saturation at the lowest temperatures may be an artefact due to joule heating. In the inset of Fig. 3a, the $T^2$- coefficient $A$ is plotted as a function of magnetic field, where the values sharply increase as the magnetic field approaches $H_{c2}(0)$. The best fit (solid line) to the field dependence of $A$ is obtained with $A(H) \propto (H-9.3)^{\alpha(=-0.5)}$. For comparison, we also show the dotted line for $\alpha = -1$, which was found for YbRh$_2$Si$_2$ where the critical behaviour was ascribed to field-induced antiferromagnetic quantum fluctuations [16]. In CeCoIn$_5$, the critical exponent is approximately -1.37 for field parallel to the c-axis [17]. We note, however, that the exponent of CeRhIn$_5$ changes when we consider the critical field $H_0$ as a fitting parameter, which was also observed in CeCoIn$_5$ [18, 19].

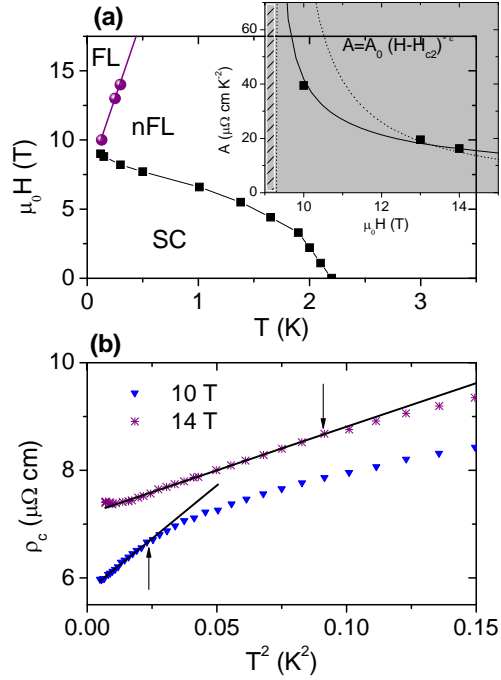

**Figure 3** (a) Magnetic field – temperature phase diagram of CeRhIn$_5$ at 2.35 GPa. Inset: resistivity $T^2$-coefficient $A$ as a function of magnetic field (see text for details). (b) Low-temperature resistivity at 10 and 14 T as a function of $T^2$. Solid lines are least-squares fits with the temperature-square dependence and arrows are the high-temperature limit above which the $T^2$-dependence is not valid.

**Table 1** Comparison of superconducting parameters for $H$ // ab-plane between CeRhIn$_5$ at 2.35 GPa and CeCoIn$_5$ at ambient pressure.

|  | CeRhIn$_5$(2.35 GPa) | CeCoIn$_5$ (1bar)[*] |
| --- | --- | --- |
| T$_c$ | 2.22±0.02 K | 2.242 K |
| -dH$_{c2}$/dT (at Tc) | -11 T/K | -30.5 T/K |
| H$_{c2}$(0) | 9.3 T | 11.73 T |
| α | 2.6 | 5.8 |

* reference [20]

Based on the resistivity measurements, we plot the *H-T* phase diagram in Fig. 3a, where the recovery of Landau-Fermi liquid behaviour in high fields strongly indicates proximity of a quantum critical point to the upper critical field $H_{c2}$. When a pressure axis is added to the *H-T* phase diagram, $H_{c2}$ is precisely the critical point where field-induced magnetism is completely suppressed, constituting the magnetic critical point [4]. In the high pressure limit, Landau-Fermi liquid responses are observed at zero field, as expected when CeRhIn$_5$ is tuned away from its quantum critical regime. When compared with the value near the quantum critical point at 10 T and 2.35 GPa (=39.42 $\mu\Omega\cdot cm\cdot K^{-2}$), the $T^2$-coefficient $A$ at 5.26 GPa (=0.046 $\mu\Omega\cdot cm\cdot K^{-2}$) is smaller at least by three orders of magnitude, which is consistent with the presence of a quantum critical point at $H_{c2}(T=0, P2)$. Table 1 compares superconducting parameters of CeRhIn$_5$ at 2.35 GPa and the field-induced quantum critical compound CeCoIn$_5$ at ambient pressure, where most of the experimental values are compatible. The Maki parameter $\alpha$ of CeCoIn$_5$ at $P=0$ is twice that of CeRhIn$_5$ but is close to $\alpha$ of CeCoIn$_5$ at 1.34 GPa [20].

## 4. Conclusion

We have reported a resistivity study of CeRhIn$_5$ at 2.35 GPa, the optimal pressure, for several magnetic fields. Similarly to the field-induced quantum critical metal CeCoIn$_5$, CeRhIn$_5$ reveals non-Fermi-liquid behaviour in the normal state, but recovers Fermi liquid properties below $T_{FL}$ for fields higher than $H_{c2}$. Approaching from higher field, the $T^2$-coefficient $A$ diverges with decreasing field, indicating that the magnetic quantum critical point is in close proximity to the upper critical field $H_{c2}(0)$.

## Acknowledgements

Work at Los Alamos was performed under the auspices of the U. S. Department of Energy/Office of Science and supported in part by the Los Alamos LDRD program. TP acknowledges a support by KOSEF grant (2009-0075786) funded by Korea government (MEST).